\documentclass[11pt,letter]{article}
\usepackage{graphicx}
\usepackage{amsmath}
\usepackage{amssymb}
\usepackage{cancel}
\usepackage{url}
\usepackage[left=1in,right=1in,top=1in,bottom=1in]{geometry}
\usepackage{setspace}
\usepackage{float}
\usepackage[super,sort&compress,comma]{natbib}
\usepackage[font={small}]{caption} 
\usepackage[labelfont=bf]{caption}
\usepackage{bm}
\usepackage[T1]{fontenc}
\usepackage{microtype}
\usepackage{chemformula}
\usepackage[version=4]{mhchem}
\usepackage{textcomp}
\usepackage{color,soul}
\usepackage{mathtools}
\usepackage{bm}
\usepackage{gensymb}
\usepackage{esvect}
\usepackage{booktabs}
\usepackage{multirow}
\usepackage{makecell}
\usepackage{physics}
\usepackage{mathrsfs}
\usepackage{booktabs}
\usepackage{longtable}
\usepackage{subcaption}

\usepackage{times}
\usepackage{url}
\usepackage{algorithm}
\usepackage{algorithmic}
\usepackage{latexsym}
\usepackage{multirow}
\usepackage{array}
\usepackage{graphicx}
\usepackage{geometry}
\usepackage{float}
\usepackage{enumerate}
\usepackage{xcolor}
\usepackage{amsmath}
\usepackage{caption}
\usepackage{diagbox}
\usepackage{chemfig}

\begin{document}

\title{\textbf{How Well Can Frontier Large Language Models Generate Structures? High Quality Prediction of Molecular Geometries with Help from Fine-Tuning}}
\author{Joseph M. Cavanagh$^{1}$, Jonathan B. Arnold$^{1}$, Giovanni Battista Alteri$^{1}$, Andrew Gritsevskiy$^{2}$\\
Teresa Head-Gordon$^{1,3,4\dagger}$}
\date{}
\maketitle

\begin{center}
$^1$Kenneth S. Pitzer Theory Center and Department of Chemistry, \\
University of California, Berkeley, CA, 94720 USA \\
$^2$RunRL, 414 Gough St, Suite 2, San Francisco CA 94102 \\
$^3$Departments of Bioengineering and Chemical and Biomolecular Engineering, \\
University of California, Berkeley, CA, 94720 USA \\
$^4$Chemical Sciences Division, Lawrence Berkeley National Laboratory, Berkeley, CA, 94720 USA \\
$\dagger$ thg@berkeley.edu
\end{center}

\begin{abstract}
\noindent
The power of Large Language Models (LLMs) has led us to investigate how they might be fine-tuned for learning the "language of molecular geometry". The fine-tuning of the LLMs using Cartesian coordinates or Z-matrices provides an extremely simple method for accurately predicting equilibrium structures and diverse sets of conformers of small organic and drug-like molecules with excellent accuracy and outperforming specialized deep learning models. While the most common representation of molecular geometries are Cartesian coordinates performs adequately, we find that the inherent invariances and relational nature of geometries represented as Z-matrices provides a better grammar for LLM adaptation. Finally, we show that enhancing an LLMs capabilities for robust prediction of small molecule geometries still retains nearly all of its pre-trained language abilities by randomly mixing in small quantities of  natural language prompt-response pairs into the fine-tuning.
\end{abstract}

\section{Introduction}
\vspace{-3mm}
The foundational nature of large language models (LLMs) have proven that they are highly adaptable for alternative language tasks from software coding to text creation and distillation.\cite{brown2020language,Anthropic2024Claude,copilot2026} The power of autoregressive, token-based LLMs has led us to investigate how they might be imbued with chemical knowledge through fine-tuning on string representations relevant to chemical tasks, including exploration of the chemical space of drug molecules\cite{yu2024llasmol, molT5, cavanaghSmileyLlamaModifyingLarge2025a}, the versatile dative bonding of transition metal complexes\cite{liuExploringTransitionMetala}, and through prediction of reaction templates and synthetic building blocks for a multi-step reaction sequence for synthesis planning of organic molecules\cite{synllama}. 

However, chemistry itself is inherently geometric through the positioning of electrons around atoms in a molecular framework; no task in chemistry is complete without the ability to model 3-dimensional structure. Previous studies have investigated the ability for language models to generate the geometry of molecules and materials in text form. These have included a demonstration that specialized, smaller language models can generate molecules, materials and protein binding sites' geometry\cite{flam-shepherdLanguageModelsCan2023}, and that LLMs in particular, specifically Llama 2\cite{touvron2023llama2openfoundation}, can be fine-tuned to generate stable inorganic materials\cite{gruver2023finetuned}. All of these studies when representing molecular geometry in text use standard file formats based on Cartesian geometries, but the reported structural root mean square deviation (RMSD) errors tend to be large when tested.

Here, we investigate the ability of LLMs to create accurate coordinates for  molecules with conformational diversity, through a suitable definition of the "language of geometry". Representational choices matter in how a pre-trained LLM can most successfully transform into an alternative language syntax, hence we consider both Cartesian coordinates as well as Z-matrix formats. We find that Z-matrices, whose built in translation and rotational invariance and relational information by construction, are a more suitable learned syntax than the Cartesian representation. Although all current frontier LLMs perform poorly in either geometry representation, when they are fine-tuned with simple prompt-response pairs they easily convert the 2D graph information of a SMILES string into reliable generator of high quality geometries across the QM9 data set. When fine-tuned on the multiple conformers of the GEOM-QM9 and GEOM-Drugs data sets, the resulting GeomLlama model realizes excellent coverage of the conformational space of molecules up to 30-40 heavy atoms with geometries that match reference low energy conformers, exceeding the performance of specialized machine learning methods. We also combine a fine-tuned LLM, SmileyLlama, to generate SMILES strings of molecules that obey Lipinski criteria to demonstrate GeomLlama's ability to predict valid, low energy conformer geometries of drug-like ligands. Finally, while fine-tuning with only structural data leads to a large degradation in natural language performance, randomly mixing in $\sim$4\% of natural language prompt-response pairs in the form of the alpaca dataset allows the model to retain nearly all of its language modeling ability while still maintaining excellent performance on geometry quality and conformational sampling.

\section{Methods}
\vspace{-2mm}

\subsection{Data Preparation and Protocols}
\vspace{-2mm}

In this work we consider multiple data sets for training and testing the ability of LLMs and special purpose geometry models to predict the Cartesian and Z-matrix structures of small organic and drug-like molecules. 
\begin{itemize}
    \item The open-source \textbf{QM9} quantum chemistry benchmark containing 133,885 organic molecules made up of carbon, oxygen, nitrogen, hydrogen, and fluorine, up to 9 heavy atoms (and up to 29 atoms with hydrogens), each with a single 3D Cartesian geometry computed using Density Functional Theory (DFT) at the B3LYP/6-31G(2df,p) level. QM9 utilizes a generic canonical SMILES representation without any stereochemistry information. We use a random split of 99\% of the data for training, and 1\% for testing.
    \item The open-source \textbf{GEOM-QM9} dataset\cite{geom} provides a conformational ensemble of the same 133,885 QM9 molecules but with structures initialized using RDKit, optimized using the semi-empirical GFN2-xTB method\cite{bannwarth_gfn2-xtbaccurate_2019}, and then subjected to extensive conformational sampling using CREST\cite{crest} to create molecular conformations. GEOM-QM9 utilizes a SMILES representation with stereochemistry information. We use a molecule-based train/validation/test split of 80/10/10 (corresponding to 1,374,737/165,204/174,162 conformations), identical to that used by Direct Molecular Conformer Generation\cite{dmcg}. 
    \item The open-source \textbf{GEOM-Drugs} dataset\cite{geom} is generated with the same CREST-based sampling procedure as GEOM-QM9, but on 430,000 molecules derived from the open-source AICures\cite{AICures2020} and  MoleculeNet\cite{moleculenet} projects. The molecules are larger than those in QM9, with an average of 25 heavy atoms per molecule, and expands the elements to also include sulfur, chlorine, and bromine. We use a molecule-based train/validation/test split of 40,000, 5,000 and 200, using the 5 lowest-energy conformers for each molecule, as pioneered by Confgf\cite{confgf} and identical to that used by Direct Molecular Conformer Generation\cite{dmcg}. This split of 200,000/25,000/1000 conformers is also used in this study.  
\end{itemize}

We convert the Cartesian coordinates of the atoms of all datasets into Z-matrices using the OpenBabel library\cite{openbabel}. Rather than the Gaussian Z-matrix format, we use the one developed by Fenske and Hall\cite{Hall1972} due to its more compact representation of the geometry. 

\subsection{Supervised Fine-Tuning}
\vspace{-2mm}

In order to steer the outputs of the pre-trained Llama-3.1-8B-Instruct model~\citep{dubey_llama_2024} for geometry generation, we use supervised fine-tuning (SFT) in which the weights of native Llama are further optimized on a dataset using the alpaca format\cite{alpaca, wang-etal-2023-self-instruct}. Given a SMILES string and corresponding geometry in the form of XYZ coordinates or a Z-matrix, we construct a prompt containing values of these two properties, with the ``correct'' completion being the named atom and its geometry in the designated molecular geometry format (Figure \ref{fig:sft_workflow}). See Supplementary Section S1 in the Supporting Information for further elaboration of the prompts used for SFT training in this study. 

\begin{figure} [H]
\centering
\includegraphics[width=0.95\textwidth]{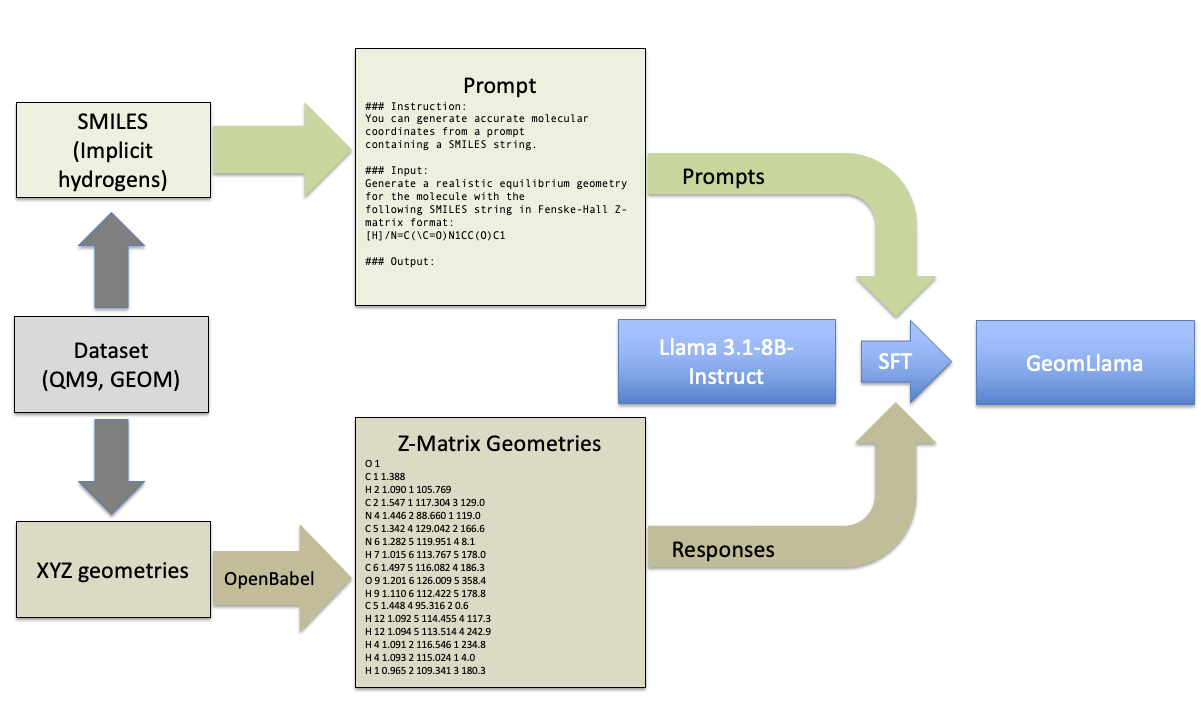}
\vspace{-2mm}
\caption{\textit{A visualization of the SFT workflow for converting LLMs to GeomLlama.} Given Qwen\cite{qwen25} and Llama~\citep{dubey_llama_2024} models, we used prompt-response pairs consisting of SMILES strings and molecular geometries to fine-tune them to do geometric conformer generation.}
\label{fig:sft_workflow}
\end{figure}

We implemented our LLM fine-tuning workflow using the axolotl framework\cite{axolotl} with a Low-Rank Adapter (LoRA) applied to all linear layers, the AdamW optimizer\cite{adamw}, a cross-entropy loss function, and a cosine learning rate scheduler. We inherit other hyperparameters, such as a LoRA rank of 32, dropout probability of 0.05, and 10 warmup steps from standard practice with axolotl in addition to previous studies in fine-tuning LLMs for chemistry\cite{synllama, cavanaghSmileyLlamaModifyingLarge2025a, liuExploringTransitionMetala}. We use the Alpaca format for instruction-tuning\cite{alpaca, wang-etal-2023-self-instruct}, a batch size of 4, and a learning rate somewhere between $2-3\times10^{-4}$. For the Llama models, we scale our model from 3B parameters to 8B parameters, training both models for 4 epochs using a batch size of 4.

During standard cross-entropy training, token probabilities are calculated using an unscaled softmax, corresponding to a temperatuer of 1.0; however, we find that varying this parameter, and the degree of nucleus sampling, can positively affect the ability of GeomLlama to sample conformers. Temperature determines the breadth of the distribution of next tokens for the LLM to be generated, such that sampling at high temperatures gives a broader and more diverse distribution, but with higher odds of incoherence. The top\_p hyperparameter, or nucleus sampling, restricts the model to only generate tokens from the smallest set whose cumulative probability sums to p. Lower values of this parameter restrict the model to only sample from high-probability tokens, while higher values of this parameter allow the model to sample from more tokens, including, perhaps, ones which are out of place. We find that a reasonable default is T=1.0, which mimics that of the training, and top\_p=0.95, which rules out most unreasonable tokens for the relatively strict format of a Cartesian or Z-matrix molecular string, such as a word in place of a number. We also find that optimal performance at inference for these hyperparameters can vary by task - such as for effective conformational sampling or when moving outside of distribution as in Sections 3.3 and 3.4.

\subsection{Frontier Large Language Models and Controls}
\vspace{-2mm}

We also assess the ability of frontier LLMs GPT-5.2\cite{openai2025gpt52}, GPT-5.4-mini\cite{openai2026gpt54mini}, Gemini-3.1-Flash\cite{google2026gemini31flashlite}, Qwen3-14B-thinking and Qwen-2.5B\cite{qwen25}, and the Llama 3 herd (Llama-3.1-8B and Llama-3.2-3B)\cite{dubey_llama_2024} to accurately guess molecular structures with a prompt used to generate a Z-matrix for a specified SMILES string from QM9:\\
\vspace{-2mm}

\noindent \textbf{Please generate the equilibrium geometry (lowest energy conformer) of the following SMILES string in Fenske-Hall z-matrix format: CC1=CCC2COC1O2. Remember that hydrogens are implicit to this SMILES string, and you will need to make them explicit in your answer. Think step-by-step about the problem before answering. Do not use an internet search to answer this question. Your final answer should include the atom names and (when applicable) distances, angles, and dihedral angles, similar to a .fh file generated with OpenBabel.}\\
\vspace{-2mm}

\noindent Hence we do not allow the models to use a code interpreter or any other tools, e.g. allowing the model to simply write RDKit code. GPT-5.2 tends to write and execute code when asked using the ChatGPT interface unless a code interpreter is expressly forbidden in a prompt as seen in Supplementary Section 1.1. For a better head-to-head comparison with our own models, we use greedy decoding.

As an additional baseline, we use some "classical" methods from RDKit\cite{Landrum2016RDKit2016_09_4} to guess the geometry of these molecules for the QM9 molecules. In particular, we use RDKit's "EmbedMolecule()" function to generate a plausible guess of the 3D coordinates.\cite{Landrum2016RDKit2016_09_4} Following this, we optimize this guessed structure with the default MMFF94 force field\cite{mmff94} from which we calculate RMSD in Sections 3.1-3.3. In Section 3.4 we use the GFN2-xTB method\cite{bannwarth_gfn2-xtbaccurate_2019} to optimize the structures from RDKit.

\subsection{Evaluation Metrics}
\vspace{-2mm}

To align with previous studies on multi-conformer predictions based on the GEOM-QM9 data set we only assess heavy-atom RMSD between predicted and generated structures, although our GeomLlama model predicts completely saturated molecular geometries. When possible we also perform graph assignment to ensure that the orderings of the atoms are correct using the rdDetermineBonds feature of RDKit\cite{Landrum2016RDKit2016_09_4}, which assigns graphs; without this, the calculated RMSD is dependent on the order in which atoms are printed. When molecular graphs are impossible to evaluate due to a structure so physically implausible that bonds cannot be assigned, we calculate RMSD based a brute-force method if there are few enough possible orderings, or the iterated Hungarian\cite{Kuhn1955}-Kabsch\cite{Kabsch1976,Kabsch1978} method if they are not.

To assess the abilities of models to sample from the space of conformers for some molecule, we use coverage (COV) and matching (MAT) scores, both calculated with 2x as many generated conformers as ground-truth conformers. The Coverage Score (COV) is defined as the proportion of true conformers $C$ with at least one generated conformer $\hat C$ which is sufficiently similar ($RMSD(C, \hat C)  < \delta$. Defining $\mathbb S_{gt}$ as the set of ground-truth conformers and $\mathbb S_{pred}$ as the set of predicted conformers, we have
\begin{equation*}
    COV(\mathbb S_{gt}, \mathbb S_{pred}) = 
    \frac{\left|\left\{ C \in \mathbb S_{gt} \;\bigg|\; \min\limits_{\hat C \in \mathbb S_{pred}} \text{RMSD}(C, \hat C) < \delta\right\}\right|}{\left|\left\{ C \in \mathbb S_{gt}\right\}\right|} \times 100\%
\end{equation*}
The Matching score (MAT) is defined as the average RMSD of the closest generated conformer to the ground truth conformers.
\begin{equation*}
    MAT(\mathbb S_{gt}, \mathbb S_{pred}) = 
    \frac{1}{\left|\left\{\mathbb S_{gt}\right\}\right|} \sum\limits_{C \in \mathbb S_{gt}} \min\limits_{\hat C \in \mathbb S_{pred}} \text{RMSD}(C, \hat C) 
\end{equation*}

\noindent In this work we use a $\delta$ cutoff set to 0.5 \AA{} for the GEOM-QM9 data sets and 1.25 \AA{} for the GEOM-Drugs data set as pused in previous studies\cite{dmcg}.


\section{Results}
\vspace{-2mm}

\subsection{Molecular Geometries of Single Equilibrium Structures from LLMs}
\vspace{-2mm}

We first investigate the ability of pre-trained LLMs to predict all-atom molecular geometries of saturated molecules from their SMILES strings\cite{smiles} for the 1339 QM9 test molecules\cite{qm9} to assess their inherent knowledge. Note that while the SMILES syntax specifies some information about molecular connectivity for the heavy atoms, it does not extend to the complete 3D geometry including stereochemistry (absent for QM9) and placement of hydrogens. Baseline methods such as RDKit provide a reference demonstrating the difficulties of generating a 3D graph from a 2D representation. As shown in Table \ref{tab:qm9}, the RDKit embedding failures can stem from difficult stereochemistry, multiple rings, and lack of convergence of the subsequent optimization to a local minimum.\cite{Landrum2016RDKit2016_09_4} 

\begin{table}[H]
    \centering
    \caption{\textit{Prediction of QM9 molecular geometries using pre-trained and fine-tuned LLMs.} For the LLM models, "Syntax (\%)" corresponds to the percentage of generations with a valid XYZ or Z-matrix format. "Atom (\%)" corresponds to the percentage of those with the correct syntax and which have the correct number of each atom element. "Graph (\%)" corresponds to the percentage of those with the correct atoms and syntax for which RDKit's RDDetermineBonds\cite{Landrum2016RDKit2016_09_4} is able to capture the molecular graph to compute all-atom RMSDs. For the RDKit\cite{Landrum2016RDKit2016_09_4} baseline we consider failures to embed to be classified as a "wrong graph". RMSD mean and medians are only reported for successful generation of a molecular geometry that pass these checks for each format. See Methods for additional details.}
    \footnotesize
    \vspace{-2mm}
    \begin{tabular}{ccccccc}
        \hline\hline
       \textbf{Format} & 
       \textbf{Model} & 
       \textbf{Syntax \%} & \textbf{Atom Count \%} & 
       \textbf{Graph \%} &
       \textbf{RMSD } & \textbf{RMSD } \\ 
        & 
        & 
       & 
       & 
       & \textbf{mean (\AA)} & \textbf{median (\AA)}  \\ \hline
        \cmidrule(lr){1-7}
        & \textbf{RDKit} & 100.00 & 100.0 & 82.22 & 1.100 & 1.168\\
        \cmidrule(lr){1-7}
        \multirow{1}{*}{\textbf{XYZ}} 
          &  \textbf{GPT-5.4-mini (low thinking)} & 100.0(0) & 22.7(8) & 12.4(6) & 1.018 & 0.978 \\
         &  \textbf{GPT-5.2} & 100.0(0) &  3.5(8) & 10.4(2) & 0.525 & 0.378 \\
         & \textbf{Gemini-3.1-Flash (preview)} & 98.3(6) & 16.7(8) & 13.1(2) & 0.862 & 0.861 \\
          & \textbf{Qwen3-14B-thinking} & 95.3(7) & 1.2(5) & 0.0(0) & -- & -- \\
         & \textbf{Llama-3.1-8B} & 98.8(1) & 0.0(0) & 0.0(0) & -- & -- \\
         & \textbf{Llama-3.2-3B} & 94.8(5) & 0.0(0) & 0.0(0) & -- & -- \\ \hline
        \cmidrule(lr){1-7}
        \multirow{4}{*}{\textbf{Z-Matrix}}
         & \textbf{GPT-5.4-mini (low thinking)} & 96.4(9) & 13.1(6) &  0.5(9) & 0.808 & 0.808 \\
         & \textbf{GPT-5.2} &  97.6(1) & 10.5(6) &  7.9(7) & 1.288 & 1.278 \\
         & \textbf{Gemini-3.1-Flash (preview)} &  92.0(8) & 15.7(3) &  6.7(0) & 1.126 & 1.228 \\
         & \textbf{Qwen3-14B-thinking} & 40.5(5) & 0.0(0) & 0.0(0) & -- & -- \\
         & \textbf{Llama-3.1-8B-Instruct} & 0.0(7) & 0.0(0) & 0.0(0) & -- & -- \\
         & \textbf{Llama-3.2-3B-Instruct} & 0.0(7) & 0.0(0) & 0.0(0) & -- & -- \\
         \cmidrule(lr){1-7}
        \multirow{1}{*}{\textbf{Fine-Tuning}} 
        & \textbf{GeomLlama-3B (XYZ)} & 100.0(0) & 99.6(3) & 88.5(3) & 0.512 & 0.322 \\
        & \textbf{GeomLlama-3B (Zmat)} & 99.5(5) & 99.7(0) & 87.4(3) & 0.331 & 0.103 \\
        & \textbf{GeomQwen-14B (Zmat)} & 99.9(3) & 99.7(8) & 90.5(6) & 0.298 & 0.102 \\
        & \textbf{GeomQwen-7B (Zmat)}  & 100.0(0) & 99.7(8) & 88.0(2) & 0.304 & 0.098 \\
        & \textbf{GeomLlama-8B (Zmat)} & 100.0(0) & 99.8(5) & 89.6(8) & \textbf{0.292} & \textbf{0.083} \\
        \hline
    \end{tabular}
\label{tab:qm9}
\end{table}

\begin{figure}[H]
\centering
\includegraphics[width=0.95\textwidth]{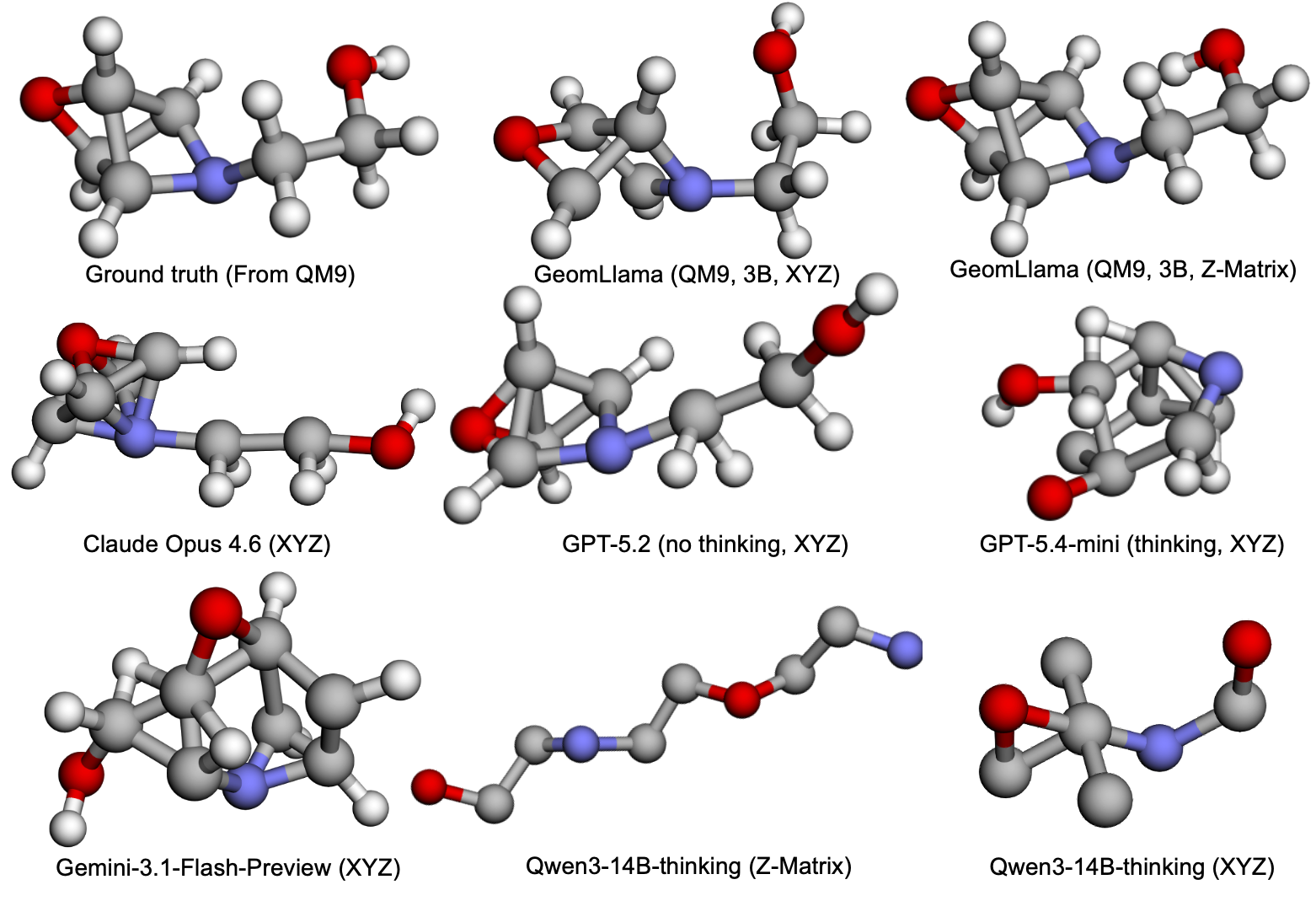}
\vspace{-2mm}
\caption{\textit{Visualization of the LLM generated geometry of a QM9 molecule with several constrained rings.} We show nine different LLMs and the geometries they create in different file formats for the molecule with SMILES string \texttt{OCCN1C2C3OC2C31}, which has several rings and no enantiomers. Each LLM is given three different attempts to generate the geometry. (a) ground-truth geometry. (b) the generated XYZ coordinates (RMSD=0.12 \AA) and (c) generated z-matrix (RMSD=0.07 \AA) from GeomLlama-3B when trained on QM9. Generated Z-matrices from (d) Claude Opus 4.6 and (e) GPT-5.2 passes atom count metric but yields a poor geometry, whereas (f) GPT-5.4-mini, (g) Gemini-3.1-Flash, and (h,i) Qwen3-14B-thinking produce molecules with incorrect atom counts.}
\label{fig:qm9}
\end{figure}

Unlike physical models used to generate 3D molecular conformations from SMILES strings, language models have far fewer enforced constraints, such that prediction of geometries as a language task will have different types of errors compared to standard benchmarks. In particular, it requires an LLM to provide a correct syntax for an XYZ file or Z-matrix format, geometry files with correct chemical elements and number of atoms expected from the SMILES string (including hydrogens), and a molecular geometry whose core connectivity must be sufficiently correct to calculate a RMSD to a reference structure. 
Table \ref{tab:qm9} shows that all frontier LLMs perform poorly on molecular geometry generation regardless of representation. 
While all pre-trained LLMs models perform better on the XYZ compared to the Z-matrix format, they also exhibit catastrophic failures for correct atom counts and thus have no legitimate graph connectivity. On the rare occasion that the pre-trained LLMs pass these basic validity checks, the structures are often still severely mis-formed (Figure \ref{fig:qm9}), such that their RMSDs that exhibit comparable performance to  RDKit are misleading. 


Table \ref{tab:qm9} and Figure \ref{fig:qm9} make clear that supervised fine-tuning of LLMs to generate molecular structures in terms of Z-matrices or XYZ coordinates leads to superior performance. To show the generality of the LLM-SFT approach, we fine-tune both open source Llama3\cite{dubey_llama_2024} and Qwen2.5\cite{qwen25} LLMs using the same SFT workflow, in which both show near perfection in syntax and atom count, and final connectivity. Their RMSDs exceed the capabilities of the RDKit software\cite{Landrum2016RDKit2016_09_4} to generate 3D chemical structures from a graph. Supplementary Table S1 uses a brute force procedure to recover from the few graph failures, and it is seen that the same conclusions hold for all models in Table \ref{tab:qm9}. 

By fine-tuning both Llama\cite{dubey_llama_2024} and Qwen\cite{qwen25}, we can also assess differences in how LLMs' tokenize numbers that might cause issues in understanding\cite{huggingfacetokenizationnumbers}. Qwen utilizes one token per digit and takes longer to generate geometry responses, while Llama3 gives nearly the same performance more efficiently by tokenizing numbers in groups of three or fewer digits\cite{dubey_llama_2024} with a shorter inference time. In regards to geometry precision for the QM9 data set in Table \ref{tab:qm9}, we train on XYZ and Z-matrices coordinates rounded "strictly" to 3 decimal places, such that "1.2001" would be rounded to "1.200" rather than "1.2", since both strings are tokenized differently; we return to the issue of precision in Sections 3.2-3.4. Additionally we find that, while there is improvement in switching from Llama 3.2-3B to Llama-3.1-8B, Qwen2.5-7B and Qwen2.5-14B perform comparably to each other, and approximately the same as observed for Llama-3.1-8B. In summary, on this most basic task of predicting single equilibrium conformations of small molecules, all GeomLlama models trained on this dataset give excellent results, but we utilize the more efficient and better Llama-3.1-8B model for the more demanding data and tasks explored in the next sections.

\subsection{Comparing GeomLlama to Deep Learning for Generation of Multiple Conformers}
\vspace{-2mm}

A number of generative deep learning approaches have been developed to predict 3D geometries for conformational ensembles of molecules based on their 2D chemical graphs, ranging from variational autoencoders\cite{dmcg,confvae} to graph\cite{cvgae} and message passing\cite{ganea2021geomol} neural networks. Here we consider the more challenging case of whether LLMs can generate conformational ensembles after SFT on the GEOM-QM9 data set, and compare them to these state-of-the-art deep learning models for molecular geometry prediction. To create a general model, we combine the GEOM-QM9 and GEOM-Drugs data sets for training, but apply the GeomLlama model separately for the GEOM-QM9 test in this section. We fine-tune the Llama-3.1-8B-instruct using the same procedure as with the QM9 dataset using the prompts show in Supplementary Section 1.2 and 1.3. 

For assessing the ability of a generative model to sample conformers from a diverse space such as that for the GEOM-QM9 and GEOM-Drugs datasets, we cannot simply use RMSD but instead use two statistical metrics: Coverage Score (COV) and matching score (MAT). The COV score measures how comprehensively a model is able to "cover" the complete distribution of actual conformers for a given molecule. The MAT score is defined as the average RMSD of the closest generated conformer to one of the ground truth conformers, and is an indicator of geometric precision across all test molecules and their conformers. These two metrics should be considered together, in which the optimal COV score would be 100\%, and the optimal MAT score would be 0.0; see Methods for more detail. 

\begin{table}[h]
    \centering
    \caption{\textit{Conformer generation on the GEOM-QM9 test set.} Performance on all-atom predictions (i.e. including hydrogens) across deep learning models and GEOMLlama trained on GEOM-QM9 and GEOM-Drugs as measured by the COV and MAT scores. In line with DMCG\cite{dmcg}, we truncate the test set to only contain molecules with between 50 and 500 conformers and evaluate RMSDs using only heavy atoms. GeomLlama models' performance is evaluated at the training temperature of 1.0 and a top\_p of 0.95. The validity of the GeomLlama results are near 100\% using the standard RDKit RDDetermineBonds to capture the molecular graph\cite{Landrum2016RDKit2016_09_4}}.
    \vspace{-2mm}
    \begin{tabular}{lcccc}
        \hline\hline
        \textbf{Method} & \textbf{COV mean} & \textbf{COV median} & \textbf{MAT mean} & \textbf{MAT median} \\ \hline
         \textbf{CVGAE\cite{cvgae}} & 0.0 & 0.0 & 1.4687 & 1.3758\\
         \textbf{GraphDG\cite{graphdg}} & 13.48 & 5.71 & 0.9511 & 0.9180\\
         \textbf{ConfVAE\cite{confvae}} & 80.18 & 85.87 & 0.3684 & 0.3776\\
         \textbf{CGCF\cite{cgcf}} & 81.48 & 86.95 & 0.3598 & 0.3684\\
         \textbf{RDKit\cite{Landrum2016RDKit2016_09_4}} & 81.61 & 85.71 & 0.2943 & 0.2472\\
         \textbf{ConfGF\cite{confgf}} & 89.21 & 95.12 & 0.2809 & 0.2837\\
         \textbf{GeoMol\cite{ganea2021geomol}} & 91.05 & 95.55 & 0.297(0) & 0.299(3)\\
        \textbf{DMCG\cite{zhu2022dmcg}} & \textbf{98.34} & \textbf{100.00} & 0.148(6) & 0.134(0)\\
        \hline
        \textbf{GeomLlama (xyz)} & 94.99 & 98.25 & 0.206(8) & 0.203(5)\\
         \textbf{GeomLlama (zmat)} & 95.76 & 98.80 & \textbf{0.133(2)} & \textbf{0.118(5)}\\
        \hline
   \end{tabular}
\label{tab:GEOMqm9}
\end{table}

Table \ref{tab:GEOMqm9} shows that many of the more recent deep learning models such as ConfGF\cite{confgf}, GeoMol\cite{ganea2021geomol}, and DMCG\cite{zhu2022dmcg} can exceed the performance of the RDKit baseline, with the DMCG method achieving the best COV and MAT scores. It is encouraging that GeomLlama is on par with DMCG in regards coverage, and generally outperforms it on MAT using the Z-matrix representation, despite being a pure LLM fine-tuned on text responses rather than a neural network adapted for molecular structure. Overall a trend emerges that the Z-matrix format is a better representation for geometry predictions, which we believe stems from the relational nature of relative coordinates versus absolute coordinates in XYZ formats, and since the latter must also contend with translational and rotational invariance. Even so, the XYZ generations are still quite accurate.

One of the most common weaknesses of Z-matrix format is that it becomes ill-defined when three atoms are co-linear, such as for molecules which contain triple bonds. OpenBabel\cite{openbabel} seeks to remediate this issue by assigning reference atoms such that no atom's position is defined with reference to the colinear atoms, which means that some of the reference distances given in the Z-matrix do not correspond to chemical bonds. When assessed on geometries for all rigid molecules within our test set, we find that whether a molecule has a triple bond or not does not seem to be detrimental to the performance of the GeomLlama models as seen in Supplementary Figure 1. In fact, our model seems to perform better on molecules with triple bonds, although that could perhaps be due to other factors, such as lacking rings, having more degrees of unsaturation, or any molecular features that would lead to a simpler overall structure.


\subsection{Geometry of Drug-like Molecules using GeomLlama}
While QM9 and GEOM-QM9 provide datasets which probe a models' abilities to generate molecular structures for a finite set of small organic molecules with 9 heavy atoms, it is far more challenging to consider the much larger chemical space of drug-like molecules. This stems from the fact that drug libraries expand the chemical elements beyond the first row of the periodic table, and are larger and more flexible, with average sizes of 25-30 heavy atoms with many more rotatable bonds. To investigate the ability of fine-tuned LLMs model to generate geometries of drug-like molecules, we test on the GEOM-Drugs\cite{geom} data set in Table \ref{tab:Drugs}. 

\begin{table}[H]
    \centering
    \caption{\textit{Conformer generation on the GEOM-Drugs test set.} Performance on all-atom predictions (i.e. including hydrogens) across deep learning models and GEOMLlama trained on GEOM-QM9 and GEOM-Drugs as measured by the COV and MAT scores. In line with DMCG\cite{dmcg}, we truncate the test set to only contain molecules with between 50 and 500 conformers and evaluate RMSDs using only heavy atoms. GeomLlama models' performance is evaluated at the training temperature of 1.0 and a top\_p of 0.95. We report performance for other temperature and top\_p values in Supplementary Figure S2.}
    \small
    \vspace{-2mm}
    \begin{tabular}{lcccccc}
        \hline\hline
        \textbf{Method} & \textbf{COV mean} & \textbf{COV median} & \textbf{MAT mean} & \textbf{MAT median} \\ 
        \hline
         \textbf{CVGAE\cite{cvgae}} & 0.00 & 0.00 & 3.0702 & 2.9937\\
        \textbf{GraphDG\cite{graphdg}} & 8.27 & 0.00 & 1.9722 & 1.9845\\
         \textbf{ConfVAE\cite{confvae}} & 53.14 & 53.98 & 1.2392 & 1.2447\\
         \textbf{CGCF\cite{cgcf}} & 53.96 & 57.06 & 1.2487 & 1.2247\\
        \textbf{RDKit\cite{Landrum2016RDKit2016_09_4}} & 60.91 & 65.70 & 1.2026 & 1.1252\\
         \textbf{ConfGF\cite{confgf}} & 62.15 & 70.93 & 1.1629 & 1.1596\\
         \textbf{GeoMol\cite{ganea2021geomol}} & 67.16 & 71.71 & 1.0875 & 1.0586\\
         \textbf{DMCG Drugs\cite{zhu2022dmcg}} & \textbf{96.52} & \textbf{100.00} & 0.7220 & 0.7161\\
         \hline
         \textbf{GeomLlama (xyz)} & 89.7(4) & 97.8(0) & 0.830(0) & 0.789(8) \\
         \textbf{GeomLlama (zmat)} & 91.0(2) & 97.33 & 0.724(7) & \textbf{0.689(3)} \\
         \textbf{GeomLlama (T-opt)} & 94.0(7) & \textbf{100.0} & \textbf{0.685(0)} & \textbf{0.642(0)} \\
        \hline
    \end{tabular}
\label{tab:Drugs}
\end{table}

We find that all deep learning models show significant degradation in the COV and MAT metrics on this more difficult data set, with the exception of DMCG which continues to excel at predicting conformational ensembles for drug-like molecules on which it is trained. Reassuringly, GeomLlama performs comparably to DMCG Drugs with just over $\sim$97\% validity, but the XYZ format has many more failures in atom count such that validity degrades to $\sim$72\%, making it no more competitive than RDKit and some of the deep learning models. 

Although GeomLlama already performs competitively on the GEOM-Drugs data in regards the Z-matrix format, its COV score is lower which we attribute to incomplete sampling. Although we could sample more conformers, here we instead take advantage of inference features that are unique to LLMs by scanning through the temperature hyperparameter value, which can be used to improve predictions for this task (see Methods). While sampling at higher temperatures can also increase the odds of invalid syntax, atom counts, or graphs, the benefit here is that drug molecules are larger and more chemically and structurally diverse. We find that the optimal performance for the hyperparameters needed for the best conformational sampling for the Z-matrix is at T=1.4 with minimal loss in validity (Supplementary Figure S2).

\subsection{Geometry of Drug-like Molecules Outside of Distribution}
\vspace{-2mm}

Due to the immense importance of drug discovery, we have previously fine-tuned Llama models to create SmileyLlama, which explores the unseen chemical space of drug-like molecules under language constraints such as obeying Lipinski's rules\cite{LIPINSKI20013} or Veber criteria\cite{veber_molecular_2002} which refers to the properties needed for oral bioavailability. Thus we not only evaluate GeomLlama performance on the test set of GEOM-Drugs, but also its performance in the molecular space in which SmileyLlama samples from under variable drug-like criteria. 

We first generated 1000 SMILES strings from SmileyLlama using the Lipinski Rule-of-Five\cite{LIPINSKI20013}, which follows physicochemical limits as multiples of 5, using the following  prompt:

\begin{itemize}
    \item \textbf{Instruction: You love and excel at generating SMILES strings of drug-like molecules.}
    \item \textbf{Input: Output a SMILES string for a drug like molecule with the following properties: <= 5 H-bond donors, <= 10 H-bond acceptors, <= 500 molecular weight, <= 5 logP}
\end{itemize}

\noindent Supplementary Figure S3 shows that under the Rule of Five SmileyLlama skews toward much larger and more complex molecules than that found for Geom-Drugs. This is an important consideration since GeomLlama is trained on GEOM-QM9 and Geom-Drugs, and hence we consider this an out-of-distribution test of the GeomLlama model. 

\begin{figure}[H]
    \centering
    \begin{subfigure}{\linewidth}
        \centering
        \includegraphics[width=0.95\linewidth]%
            {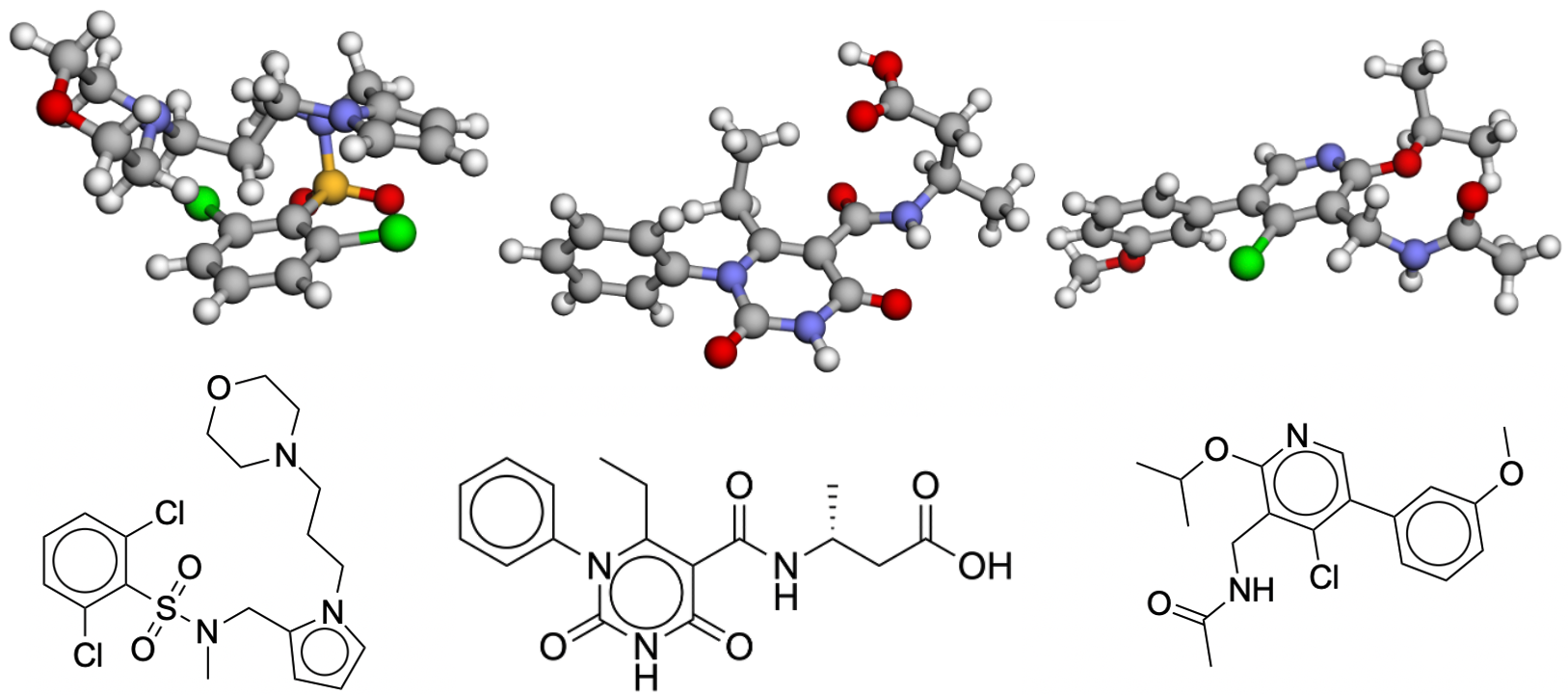}
        \phantomsubcaption\label{fig:plot_a}
    \end{subfigure}\\[-0.5ex]
    \vspace{2mm}
    \begin{subfigure}{\linewidth}
        \centering
        \includegraphics[width=0.95\linewidth]%
            {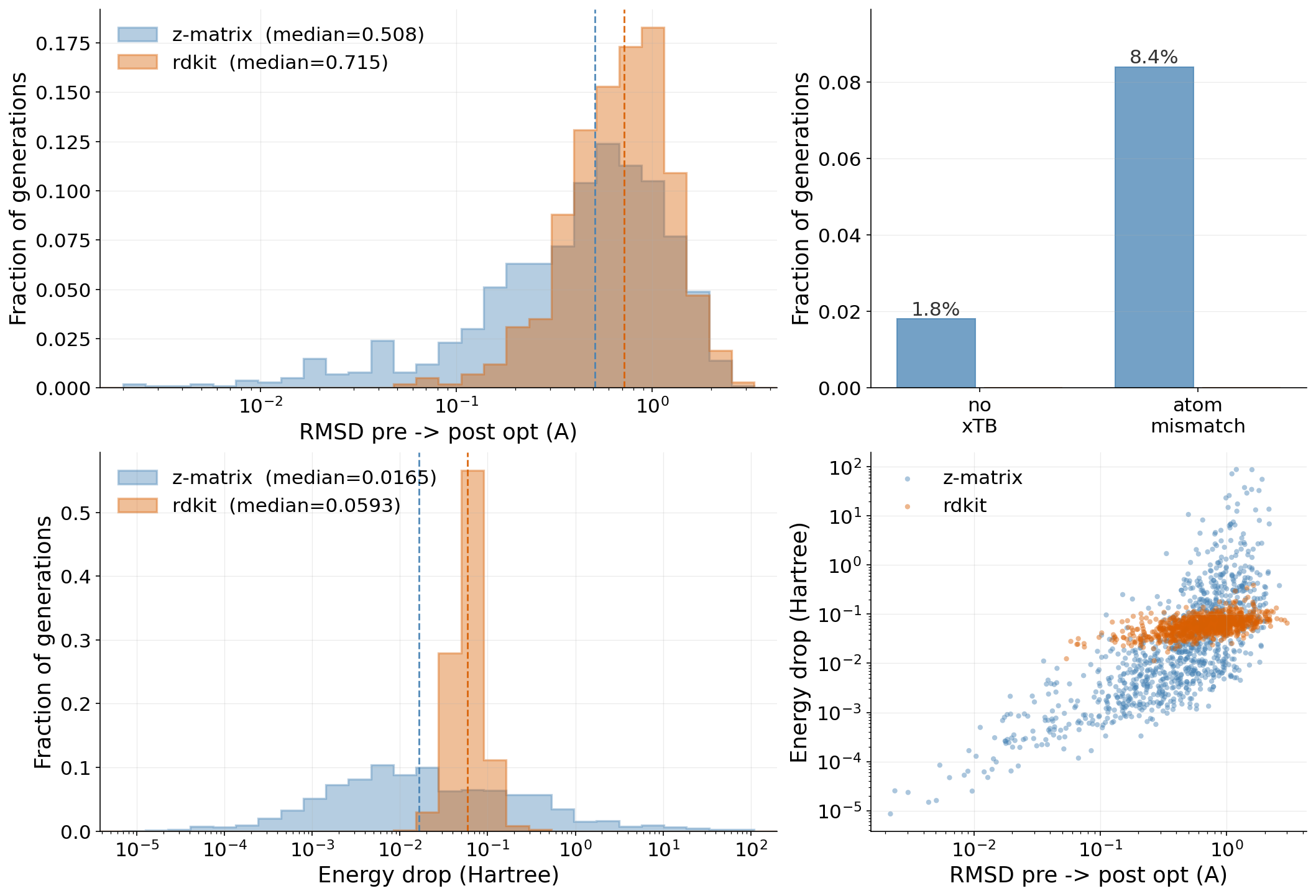}
        \phantomsubcaption\label{fig:plot_b}
    \end{subfigure}\\[-0.5ex]
    \caption{\textit{SmileyLlama drug-like molecules prompted with Lipinski's Rule of Four.} \textbf{(a)} Representative geometries using GeomLlama trained with Z-matrices. \textbf{(b)} Geometry differences between SmileyLlama structures generated by GeomLlama Rule-of-Four and RDKit's \texttt{EmbedMolecules}, compared against their GFN2-xTB-optimized geometries. It is seen that GeomLlama has $\sim$8.4\% validity failures and a small number of optimization failures, wherein GFN2-xTB fails to converge. \textbf{(c)} change in energy after optimizing GeomLlama's structures as well as RDKit's structures. (d) Change in geometry versus change in energy upon optimization with GFN2-xTB. The positive correlation in energy drop and RMSD indicate that the starting structures from GeomLlama are closer to the local minimum compared to those of RDkit on the GFN2-xTB energy surface.}
    \label{fig:both_plots}
\end{figure}

We also consider a second prompt based on the Rule of Four:

\begin{itemize}
    \item \textbf{Instruction: You love and excel at generating SMILES strings of drug-like molecules.}
    \item \textbf{Input: Output a SMILES string for a drug like molecule with the following properties: <= 4 H-bond donors, <= 4 H-bond acceptors, <= 400 molecular weight, <= 4 logP}
\end{itemize}

\noindent which then responds with 1000 SMILES strings, resulting in properties that now show much better overlap with the GEOM-Drugs data set as also seen in Supplementary Figure S3, and thus serves as an in-distribution sample to test GeomLlama. Example molecules generated by SmileyLlama followed by GeomLlama optimization for Rule of Four and Rule of Five are shown in Figure \ref{fig:both_plots} (top).

Starting with the in-distribution test set on the Rule-of-Four, we use GeomLlama to generate 10 geometries for each of the 1000 SMILES strings, selecting the first with the correct atom count and no interatomic distance under 0.7 \AA{} to account for steric clashes. 
Because there are no ground truth structures as found in a benchmark library like GEOM-Drugs, we instead test how closely these generated starting structures are to the local minimum found by an GFN2-xTB\cite{bannwarth_gfn2-xtbaccurate_2019} optimization, both in terms of geometry (RMSD) and GFN2-xTB energy, and compare to RDKit embedded geometries and energies that are also optimized with GFN2-xTB. We find that 84 of the 1000 Rule-of-Four molecules fail to generate a valid Z-matrix using GEOMLlama due to a wrong atom count, and of the remaining 916 molecules a little under 2\% fail to 

\noindent converge using GFN2-xTB. But for the remaining $\sim$90\% of molecules we generate geometries which are significantly closer to the xTB minimum compared to RDKit structures in terms of RMSD and energy (Figure \ref{fig:both_plots} (bottom)). Finally, Supplementary Figure S4 shows the same analysis for the out-of-distribution case of SmileyLlama Rule-of-Five results. Here we see that atom mismatches are only 10\% and optimization failures rise to 4\%, but overall geometries are superior to RDKit.

\subsection{Natural language modeling outside of conformer generation}
One question that naturally arises is whether training on conformer generation atrophies the LLM's abilities in natural language processing, and, if so, to what extent it can be mitigated. We find that the simplest procedure for converting an LLM into a competent conformer generator, by fine-tuning purely on this type of data, also diminishes its ability to answer questions as shown in the prompt responses in Supplementary Section 1.4.


\begin{figure} [H]
\centering
\includegraphics[width=0.95\textwidth]{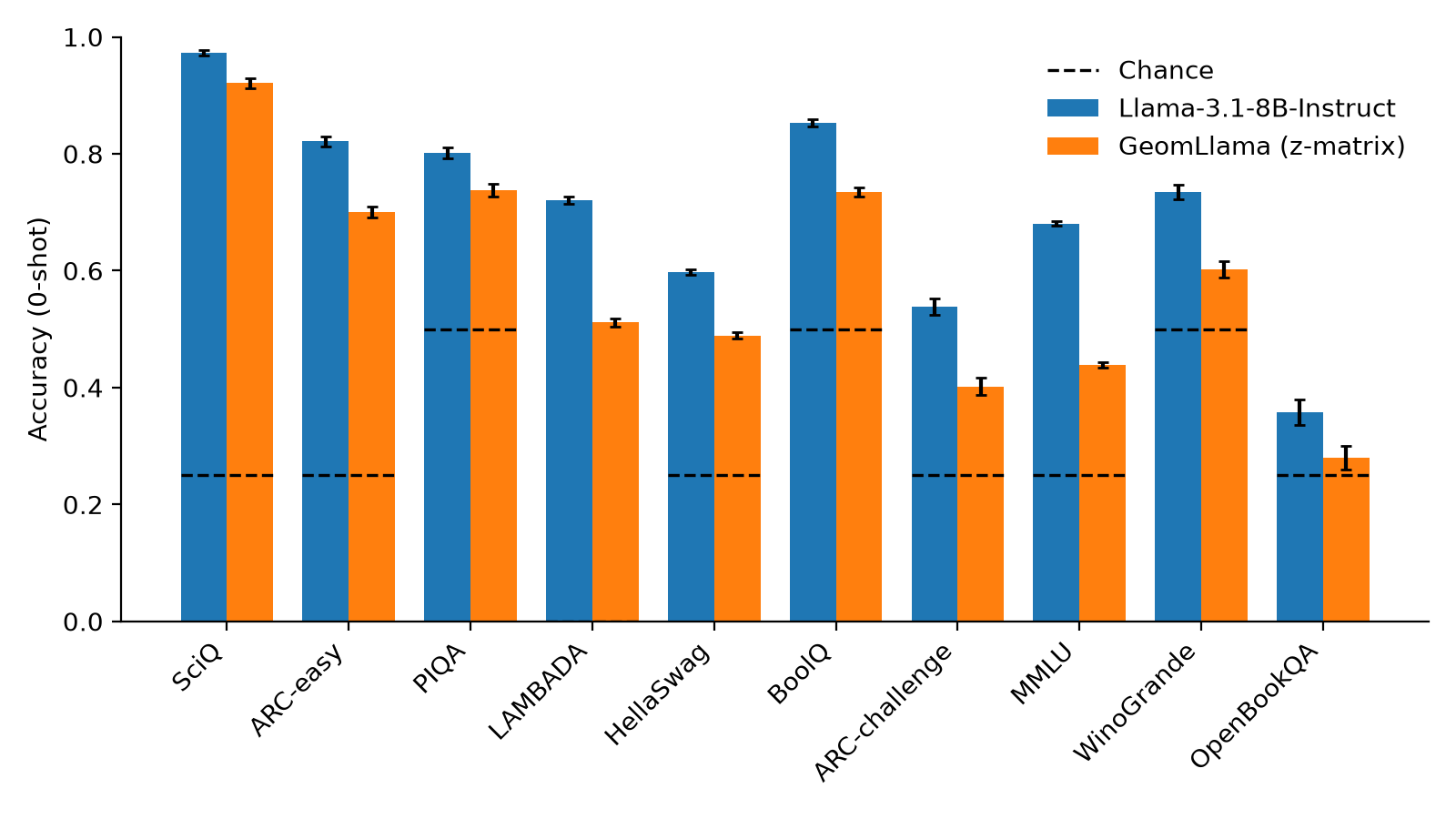}
\vspace{-2mm}
\caption{\textit{Performance gap across various benchmarks\cite{eval-harness} for pre-trained Llama-3.1-8B vs. GeomLlama (z-matrix with pseudoreherasal.} All benchmarks are scored between 0 and 1, with 1 representing perfect performance. Fine-tuning causes performance to deteriorate across all datasets, although this is less pronounced in tests of fact recall (SciQ, ARC-Easy, BoolQ)\cite{sciq, arc2, boolq}, general language processing ability (LAMBADA, HellaSwag)\cite{lambada, hellaswag}, or common-sense QA (PIQA)\cite{piqa}, than it is in tests of more complex knowledge synthesis (ARC-challenge, MMLU, WinoGrande, OpenBookQA)\cite{arc2, mmlu, winogrande,OpenBookQA}.}
\label{fig:mmlu}
\end{figure}

We sought to mitigate this issue using three primary strategies. The first was to reduce the LoRA rank\cite{biderman2024loralearnsforgets}, to reduce forgetting at the possible expense of performance of GeomLlama, and found that this does not substantially affect our model's performance on MMLU/Wikitext perplexity, although it did deteriorate performance in conformer generation. Second, we used RDKit \cite{Landrum2016RDKit2016_09_4} to generate a long list of molecular properties in natural language that were added to the prompt during training. We found that this also did not mitigate loss of natural language processing ability. Finally, we randomly shuffled the alpaca dataset into our training set, which is an instruction-tuning dataset consisting of multiple choice question-answer pairs. This strategy of mitigating catastrophic forgetting by introducing samples of data similar to the original training set of our model is called Pseudorehearsal\cite{ROBINS01061995}. We found that, despite alpaca containing less than 4\% of the number of training examples as GEOM-QM9, training Llama-3.1-8B-instruct on geometry generation with a mix of the GEOM-QM9 and alpaca datasets let the model keep its ability to process natural language, and gain the ability to generate conformers as well as a model trained purely on GEOM-QM9. (Supplementary Table S2). To put this into perspective, the Z-matrix trained GeomLlama's natural language abilities are in the neighborhood of the comparably sized Llama 2-7B model\cite{touvron2023llama2openfoundation}. \textbf{to be updated with better results}

\section{Discussion and Conclusion}
\vspace{-2mm}

The advancing foundational power of the large language models of today are becoming increasingly general-purpose, as shown by the recently released Agent's Last Exam benchmark\cite{sun2026agentsexam}. But as of today, out-of-the-box LLMs are unable to generate molecular geometries reliably given a SMILES string as we have shown here. Rather than trying to find whether LLMs can generate stable structures from a blank prompt like the previous studies that we build on\cite{}, we use the ability of an LLM to respond to information in its prompt to convert it into a conformer generator, which can easily and natively convert the 2d graph information contained in a SMILES into a high quality structure. Although GeomLlama is trained to only predict geometries, it implicitly finds stable low energy conformations of unseen small organic and drug-like molecules up to 30-40 heavy atoms, including the predictions of the saturating hydrogen positions. In doing this, we find that the Z-matrix representation is broadly superior to Cartesian coordinates, likely due to its inherent invariance properties and relational nature of common chemical motifs embodied in bond lengths and angles which can be effectively memorized and applied. 

Even so, there are still limitations to and tradeoffs within the GeomLlama framework- primarily its small but still non-negligible failure rates on valid atom counts and structure file formats as molecules get larger. These validity failure modes are peculiar to language models, rooted at their origin in learning a correct syntax, and unlike the failure modes of RDKit or specialized deep learning models for geometry. While some improvements to GeomLlama will come from larger foundation models and standard hyperparameter tuning, at the same time overcoming these limitations would also be found in the unique solutions possible with LLMs, such as better prompt engineering and the use of direct preference optimization to improve adherence to the prompt. Furthermore, GeomLlama still struggles in data-poor regimes, for instance in the task of generating geometries beyond the 30-40 heavy atoms demonstrated in this work. Hence expanding data sets to larger and more complex molecules will continue to be a priority for the field.

Finally, we showed that adding this new ability to speak the language of molecular geometry can be done with a Low-Rank Adapter, which can be cheaply added and taken away from LLMs during inference--- effectively meaning that this ability to generate structures can be given to the model when generating a geometry and taken away with only a very small inference cost, when using contemporary inference frameworks such as vLLM\cite{vllm}. We additionally find that, while fine-tuning with only this data leads to a large degradation in natural language retention, randomly mixing in natural language prompt-response pairs in the form of the alpaca dataset allows the model to retain much of its language modeling ability. Establishing the capability of an LLM to generate accurate molecular structures without taking away from its other inherent  knowledge bodes well for AI to revolutionize the field of chemistry. 

\section*{Data Availability}
\vspace{-2mm}

\noindent All data generated in this study can be found in \texttt{https://github.com/THGLab/GeomLlama}. The data included here contains all SMILES strings, Z-matrices, and XYZ coordinates used for training, validation, and testing, the dataset containing prompts and responses used for SFT to create each GeomLlama model from Llama. Source data for all Figures is available with this manuscript. 

\section*{Code Availability}
\vspace{-2mm}

\noindent All code generated in this study can be found in \texttt{https://github.com/THGLab/GeomLlama}.. 

\section*{Acknowledgments}
\vspace{-2mm}

\noindent We thank Dr. Yunsheng Liu and Yuchen Shi for helpful discussions. This work was supported in part by the National Institute of Allergy and Infectious Disease grant U19-AI171954 for the drug molecule application. We thank the CPIMS program, Office of Science, Office of Basic Energy Sciences, Chemical Sciences Division of the U.S. Department of Energy under Contract DE-AC02-05CH11231 for support of the machine learning. This work used Delta AI at NCSA through allocation CHE250012 from the Advanced Cyberinfrastructure Coordination Ecosystem: Services \& Support (ACCESS) program, which is supported by U.S. National Science Foundation grants \#2138259, \#2138286, \#2138307, \#2137603, and \#2138296\cite{access}. 

\section*{Author Contributions Statement}
\vspace{-2mm}

\noindent J.M.C. and T.H.G. defined goals and designed the project. J.M.C., J.B.A, A.G., and G.B.A. performed the calculations. T.H.-G. carefully ensured that all results were completely described and analyzed correctly. J.M.C. and T.H.G. wrote and edited the paper. All authors discussed the results and made comments to the manuscript.

\section*{Competing Interests Statement}
\vspace{-2mm}

\noindent The authors declare no competing interests.

\bibliography{library}
\end{document}


\title{\textbf{How Well Can Frontier Large Language Models Generate Structures? High Quality Prediction of Molecular Geometries with Help from Fine-Tuning}}
\author{Joseph M. Cavanagh$^{1}$, Jonathan B. Arnold$^{1}$, Giovanni Battista Alteri$^{1}$,\\
Teresa Head-Gordon$^{1,2,3\dagger}$}
\date{}
\maketitle

\begin{center}
$^1$Kenneth S. Pitzer Theory Center and Department of Chemistry, \\
University of California, Berkeley, CA, 94720 USA \\
$^2$Departments of Bioengineering and Chemical and Biomolecular Engineering, \\
University of California, Berkeley, CA, 94720 USA \\
$^3$Chemical Sciences Division, Lawrence Berkeley National Laboratory, Berkeley, CA, 94720 USA \\
$\dagger$ thg@berkeley.edu
\end{center}

\section{Prompts}
Below are some example prompts used for frontier and fine-tuned LLMs for generating molecular geometries in both xyz and z-matrix formats.

\subsection{Frontier model sample prompt-response pair}
Below is an example of a prompt and response from our study of GPT-5.2's ability to process this:
\begin{tcolorbox}[title=System Prompt,breakable]
\begin{verbatim}
You are a computational chemistry assistant. When given a SMILES 
string, you generate plausible 3D molecular geometries in XYZ 
format purely from your training knowledge.

Rules:
- Do NOT write or execute any code.
- Do NOT use RDKit, Open Babel, or any cheminformatics toolkit.
- Do NOT use any external tools or calculators.
- Respond ONLY with the XYZ coordinate block. No explanation, no 
commentary.

The format is:
<element> <x> <y> <z>
...

Coordinates should be in Angstroms with reasonable bond lengths and 
angles.
\end{verbatim}
\end{tcolorbox}
\begin{tcolorbox}[title=User Prompt,breakable]
\begin{verbatim}
N#CC12CC1CO2
\end{verbatim}
\end{tcolorbox}
\begin{tcolorbox}[title=Response,breakable]
\begin{verbatim}
C 0.0000 0.0000 0.0000
O 1.4200 0.0000 0.0000
C 2.8500 0.0000 0.0000
C 3.6200 1.2600 0.0000
O 3.0500 2.4700 0.0000
C 5.1400 1.2600 0.0000
C 6.3400 1.2600 0.0000
C 7.5400 1.2600 0.0000
C 8.8800 1.2600 0.0000
O 10.0800 1.2600 0.0000
H -0.5400 0.9400 0.0000
H -0.5400 -0.4700 0.8100
H -0.5400 -0.4700 -0.8100
H 3.2100 -0.9400 0.0000
H 3.0600 1.2600 0.9400
H 3.0600 1.2600 -0.9400
H 5.5200 2.2000 0.0000
H 5.5200 0.7900 0.8100
H 5.5200 0.7900 -0.8100
H 8.9800 2.3500 0.0000
\end{verbatim}
\end{tcolorbox}

\subsection{Cartesian (xyz) format}
\begin{tcolorbox}[title=System Prompt,breakable]
\begin{verbatim}
You can generate accurate molecular coordinates from a prompt
containing a SMILES string.
\end{verbatim}
\end{tcolorbox}
\begin{tcolorbox}[title=User Prompt,breakable]
\begin{verbatim}
Generate a realistic equilibrium geometry for the molecule with the 
following SMILES string in xyz format: 
[H]/N=C(\C=O)N1CC(O)C1
\end{verbatim}
\end{tcolorbox}
\begin{tcolorbox}[title=Response,breakable]
\begin{verbatim}
O -3.051362 -0.375362 0.442857
C -2.005681 -0.107853 -0.430127
H -2.439175 -0.138709 -1.429971
C -1.145204 1.146061 -0.143971
N 0.005051 0.281476 0.002864
C 1.307008 0.605823 0.042510
N 1.658384 1.837521 0.098974
H 2.662754 1.974249 0.157370
C 2.270829 -0.540149 0.057400
O 1.981710 -1.706057 0.064390
H 3.328597 -0.203641 0.069164
C -0.713853 -0.945350 -0.268175
H -0.377418 -1.478919 -1.159328
H -0.722179 -1.641677 0.575225
H -1.047698 1.866646 -0.957569
H -1.399839 1.681898 0.774132
H -2.723068 -0.356127 1.350340
\end{verbatim}
\end{tcolorbox}
\subsection{z-matrix format}
\begin{tcolorbox}[title=System Prompt,breakable]
\begin{verbatim}
You can generate accurate molecular coordinates from a prompt
containing a SMILES string.
\end{verbatim}
\end{tcolorbox}
\begin{tcolorbox}[title=User Prompt,breakable]
\begin{verbatim}
Generate a realistic equilibrium geometry for the molecule with the 
following SMILES string in Fenske-Hall Z-matrix format: 
[H]/N=C(\C=O)N1CC(O)C1
\end{verbatim}
\end{tcolorbox}
\begin{tcolorbox}[title=Response,breakable]
\begin{verbatim}
O 1
C 1 1.388
H 2 1.090 1 105.769
C 2 1.547 1 117.304 3 129.0
N 4 1.446 2 88.660 1 119.0
C 5 1.342 4 129.042 2 166.6
N 6 1.282 5 119.951 4 8.1
H 7 1.015 6 113.767 5 178.0
C 6 1.497 5 116.082 4 186.3
O 9 1.201 6 126.009 5 358.4
H 9 1.110 6 112.422 5 178.8
C 5 1.448 4 95.316 2 0.6
H 12 1.092 5 114.455 4 117.3
H 12 1.094 5 113.514 4 242.9
H 4 1.091 2 116.546 1 234.8
H 4 1.093 2 115.024 1 4.0
H 1 0.965 2 109.341 3 180.3
\end{verbatim}
\end{tcolorbox}

\subsection{Language ability of GEOM-Llama with and without pseudorehearsal}
\begin{tcolorbox}[title=System Prompt,breakable]
\begin{verbatim}
You are a brilliant and helpful AI assisstant who answers any 
question given as input truthfully, accurately, and precisely.
\end{verbatim}
\end{tcolorbox}
\begin{tcolorbox}[title=User Prompt,breakable]
\begin{verbatim}
What is 1+1?
\end{verbatim}
\end{tcolorbox}
\begin{tcolorbox}[title=Response (Z-matrix),breakable]
\begin{verbatim}
O 1
C 1 1.200
C 2 1.439 1 125.979
C 2 2.639 1 125.981 3 0.0
\end{verbatim}
\end{tcolorbox}
\begin{tcolorbox}[title=Response (XYZ),breakable]
\begin{verbatim}
1.000 0.000 0.000
\end{verbatim}
\end{tcolorbox}
\begin{tcolorbox}[title=Response (Z-matrix with Alpaca Pseudorehearsal),breakable]
\begin{verbatim}
2
\end{verbatim}
\end{tcolorbox}

\begin{tcolorbox}[title=Text and completion (Z-matrix),breakable]
\begin{Verbatim}[commandchars=\\\{\}]
They say that a bird in the hand is worth two in the \textcolor{red}{201.0}
\textcolor{red}{C 1}
\textcolor{red}{C 1 1.505}
\textcolor{red}{C 2 1.329}
\textcolor{red}{C 3 1.463 1 123.000}
\textcolor{red}{O 4 1.204 3 124.620 1 180.0}
\textcolor{red}{C 4 1.506 3 112.677 1 0.0}
\textcolor{red}{C 2 1.475 1 121.600 3 180.0}
\textcolor{red}{N 6 1.448 2 121.862 1 360.0}
\end{Verbatim}
\end{tcolorbox}
\begin{tcolorbox}[title=Text and completion (XYZ),breakable]
\begin{Verbatim}[commandchars=\\\{\}]
They say that a bird in the hand is worth two in the \textcolor{red}{0.000}
\textcolor{red}{A 2.678 0.748 0.000}
\textcolor{red}{C 1.580 -0.000 -0.749}
\textcolor{red}{C 0.260 0.000 0.000}
\textcolor{red}{C -0.900 -0.000 -0.990}
\textcolor{red}{C -2.242 0.000 -0.292}
\textcolor{red}{O -2.404 1.183 0.456}
\textcolor{red}{C -2.242 -1.183 0.456}
\textcolor{red}{N 1.580 -0.000 0.749}
\end{Verbatim}
\end{tcolorbox}
\begin{tcolorbox}[title=Text and completion (Z-matrix with Alpaca Pseudorehearsal),breakable]
\begin{Verbatim}[commandchars=\\\{\}]
They say that a bird in the hand is worth two in the\textcolor{red}{ bush.}
\textcolor{red}{But when it comes to the internet, it's not just about the birds}
\textcolor{red}{you have, it's also about the ones you can catch with your net.}
\textcolor{red}{So, if you're looking to get the most out of your online presence, }
\textcolor{red}{you should be aware of the different ways you can use the internet }
\textcolor{red}{to your advantage. Here are some tips to help you make the most of }
\textcolor{red}{your online experience: 1. Keep your profile and social media accounts}
\textcolor{red}{up to date. This will help you build a strong online presence and }
\textcolor{red}{make it easier for people to find you. 2. Use the right keywords}
\end{Verbatim}
\end{tcolorbox}

\section{Tables}

\begin{table}[H]
    \centering
    \caption{\textit{Prediction of molecular geometries from QM9 using pre-trained and fine-tuned LLMs.} The RDKit\cite{Landrum2016RDKit2016_09_4} baseline embeds each SMILES string and then optimizes with MMFF94\cite{}; for this baseline, we consider failures to embed to be classified as a "wrong syntax".  For other models, "Syntax (\%)" corresponds to the percentage of generations with a valid a z-matrix or xyz format; "Atom (\%)" corresponds to the percentage which have the correct number of each atom element. We use RDKit's RDDetermineBonds\cite{Landrum2016RDKit2016_09_4} to capture the molecular graph, and if that fails a brute force approach to determine the ordering of the heavy atoms which minimizes RMSD, followed by iterative Hungarian algorithm + Kabsch algorithm with hydrogens. RMSD mean and medians are only reported for successful generation of a molecular geometry that pass the syntax and atom count checks for each format.}
    \footnotesize
    \vspace{-2mm}
    \begin{tabular}{cccccc}
        \hline\hline
       \textbf{Format} & 
       \textbf{Model} & 
       \textbf{Syntax \%} & 
       \textbf{Atom Count \%} & 
       \textbf{RMSD mean (\AA) } & \textbf{RMSD median (\AA)} \\ 
        \cmidrule(lr){1-6}
        & \textbf{RDKit} & 100.0 & 100.0 & 1.100(3) & 1.168(1)\\ 
        \cmidrule(lr){1-6}
        \multirow{1}{*}{\textbf{XYZ}} 
          &  \textbf{GPT-5.4-mini (low thinking)} & 100.0(0) & 22.7(8) & 1.162 & 1.176\\
         &  \textbf{GPT-5.2} & 100.0(0) &  3.5(8) & 1.106 & 1.174\\
         & \textbf{Gemini-3.1-Flash (preview)} & 98.3(6) & 16.7(8) & 1.114 & 1.122\\
          & \textbf{Qwen3-14B-thinking} & 95.3(7) & 1.2(5) & 1.351 & 1.364\\
         & \textbf{Llama-3.1-8B} & 98.8(1) & 0.0(0) & -- & --\\
         & \textbf{Llama-3.2-3B} & 94.8(5) & 0.0(0) & -- & --\\
        \cmidrule(lr){1-6}
        \multirow{1}{*}{\textbf{Z-Matrix}}
         & \textbf{GPT-5.4-mini (low thinking)} & 96.4(9) & 13.1(6) & 1.379 & 1.375\\
         & \textbf{GPT-5.2} &  97.6(1) & 10.5(6) & 1.350 & 1.346\\
         & \textbf{Gemini-3.1-Flash (preview)} &  92.0(8) & 15.7(3) & 1.349 & 1.359\\
         & \textbf{Qwen3-14B-thinking} & 40.5(5) & 0.0(0) & -- & -- \\
         & \textbf{Llama-3.1-8B-Instruct} & 0.0(7) & 0.0(0) & -- & -- \\
         & \textbf{Llama-3.2-3B-Instruct} & 0.0(7) & 0.0(0) & -- & -- \\
         \cmidrule(lr){1-6}
        \multirow{1}{*}{\textbf{Fine-Tuning}} 
        & \textbf{GeomQwen-14B (Z-matrix)} & 99.9(3) & 99.7(8) & 0.320 & 0.112\\
        & \textbf{GeomQwen-7B (Z-matrix)}  & 100.0(0) & 99.7(8) & 0.336 & 0.120 \\
        & \textbf{GeomLlama-8B (Z-matrix)} & 100.0(0) & 99.8(5) & 0.367 & 0.131 \\
        & \textbf{GeomLlama-3B (Z-matrix)} & 99.5(5) & 99.7(0) & 0.320 & 0.095\\
        & \textbf{GeomLlama-3B (XYZ)} & 100.0(0) & 99.6(3) & 0.522 &  0.347\\
        \hline
    \end{tabular}
\label{tab:qm9}
\end{table}

\begin{table}[H] 
\begin{center}
 \caption{\textit{Standard language model evaluation metrics for the GeomLlama model trained with Pseudorehearsal}. Results are shown for training with GEOM-QM9 and GEOM-Drugs using the Z-matrix format combined with 4\% of the alpaca dataset}
  \begin{tabular}{lccc}
  \hline \hline
  \textbf{Model} & Llama-3.1-8B-Instruct & GeomLlama & Chance \\
  && (\texttt{z-matrix}, pseudorehearsal) & \\
  \hline \hline
Wikitext (bits per byte) ($\downarrow$) & 0.5818 & 0.9156 & \\
\hline
MMLU (0-shot) ($\uparrow$)  & 0.6807 ± 0.0037 & 0.4388 ± 0.0041 & \\
\quad Humanities            & 0.6436 ± 0.0067 & 0.4029 ± 0.0069 & \\
\quad Social Sciences       & 0.7702 ± 0.0075 & 0.5190 ± 0.0089 & \\
\quad STEM                  & 0.5858 ± 0.0074 & 0.3666 ± 0.0085 & \\
\quad Other                 & 0.7444 ± 0.0084 & 0.4870 ± 0.0088 &\\
\hline
  LAMBADA (0-shot) ($\uparrow$)       & 0.7206 ± 0.0063 & 0.5119 ± 0.0070 & --- \\
  HellaSwag (0-shot) ($\uparrow$)     & 0.5979 ± 0.0049 & 0.4892 ± 0.0050 & 0.25 \\
  PIQA (0-shot) ($\uparrow$)          & 0.8020 ± 0.0093 & 0.7378 ± 0.0103 & 0.50 \\
  WinoGrande (0-shot) ($\uparrow$)    & 0.7340 ± 0.0124 & 0.6022 ± 0.0138 & 0.50 \\
  ARC-easy (0-shot) ($\uparrow$)      & 0.8211 ± 0.0079 & 0.6999 ± 0.0094 & 0.25 \\
  ARC-challenge (0-shot) ($\uparrow$) & 0.5384 ± 0.0146 & 0.4019 ± 0.0143 & 0.25 \\
  BoolQ (0-shot) ($\uparrow$)         & 0.8535 ± 0.0062 & 0.7339 ± 0.0077 & 0.50 \\
  OpenBookQA (0-shot) ($\uparrow$)    & 0.3580 ± 0.0215 & 0.2800 ± 0.0201 & 0.25 \\
  SciQ (0-shot) ($\uparrow$)          & 0.9730 ± 0.0051 & 0.9210 ± 0.0085 & 0.25 \\
  \hline \hline
  \end{tabular}
  \end{center}
  \end{table}

\section{Figures}

\begin{figure} [H]
\centering
\includegraphics[width=0.7\textwidth]{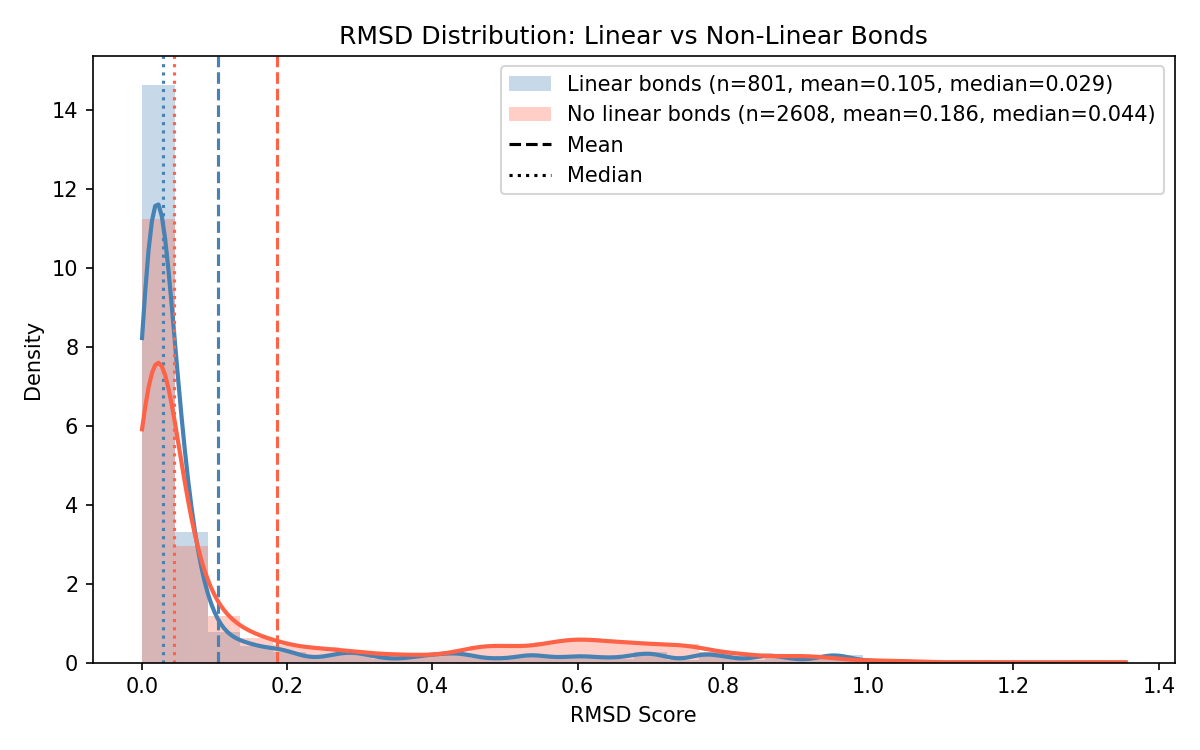}
\vspace{-2mm}
\caption{\textit{Comparison of GeomLlama model Z-matrix performance for molecules with a triple bond compared to those without.} GeomLlama has no significant problems with molecules with linear bonds.}
\label{fig:geom-temp-dependence}
\end{figure}

\begin{figure} [H]
\centering
\includegraphics[width=0.99\textwidth]{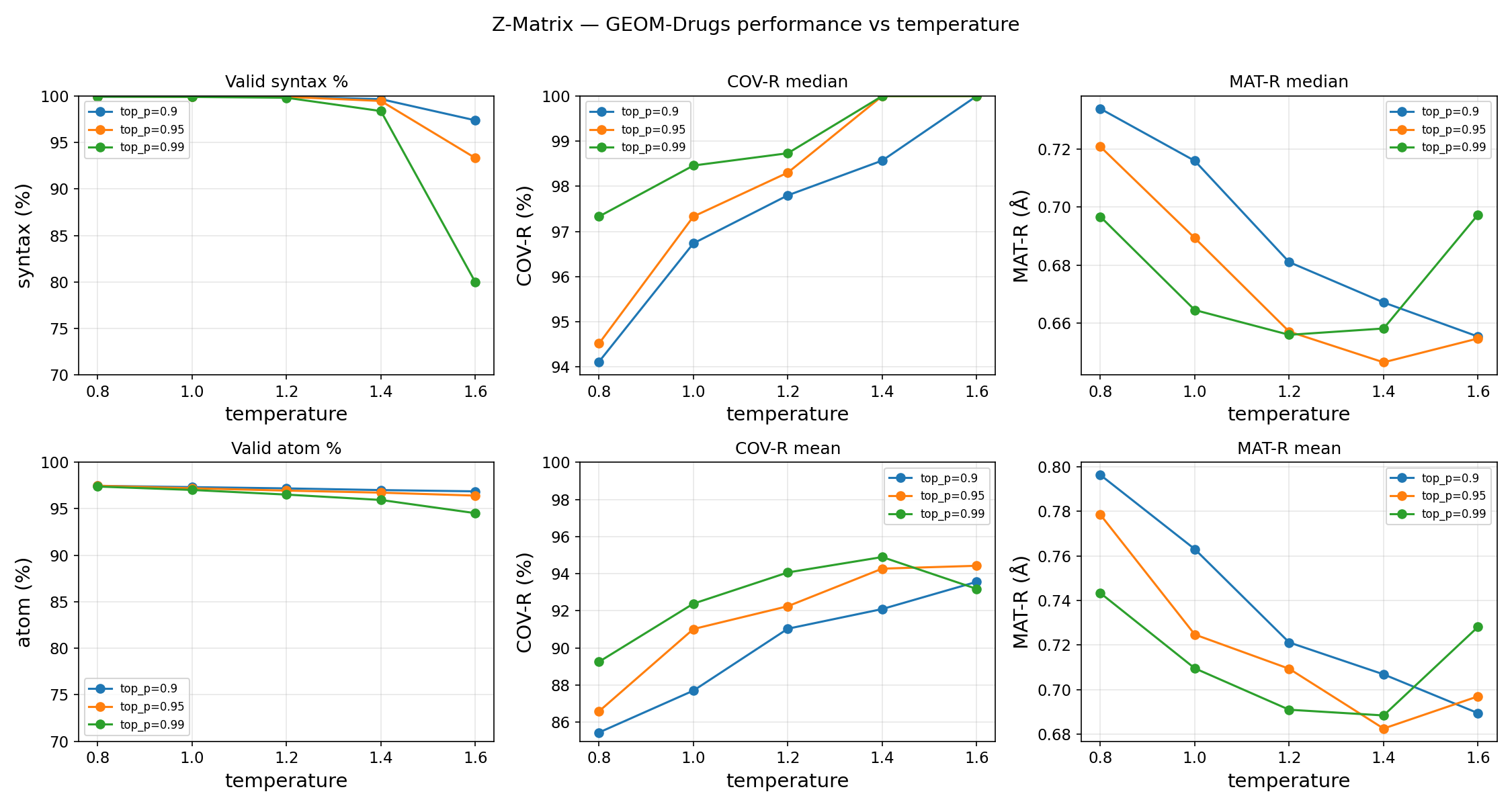}
\includegraphics[width=0.99\textwidth]{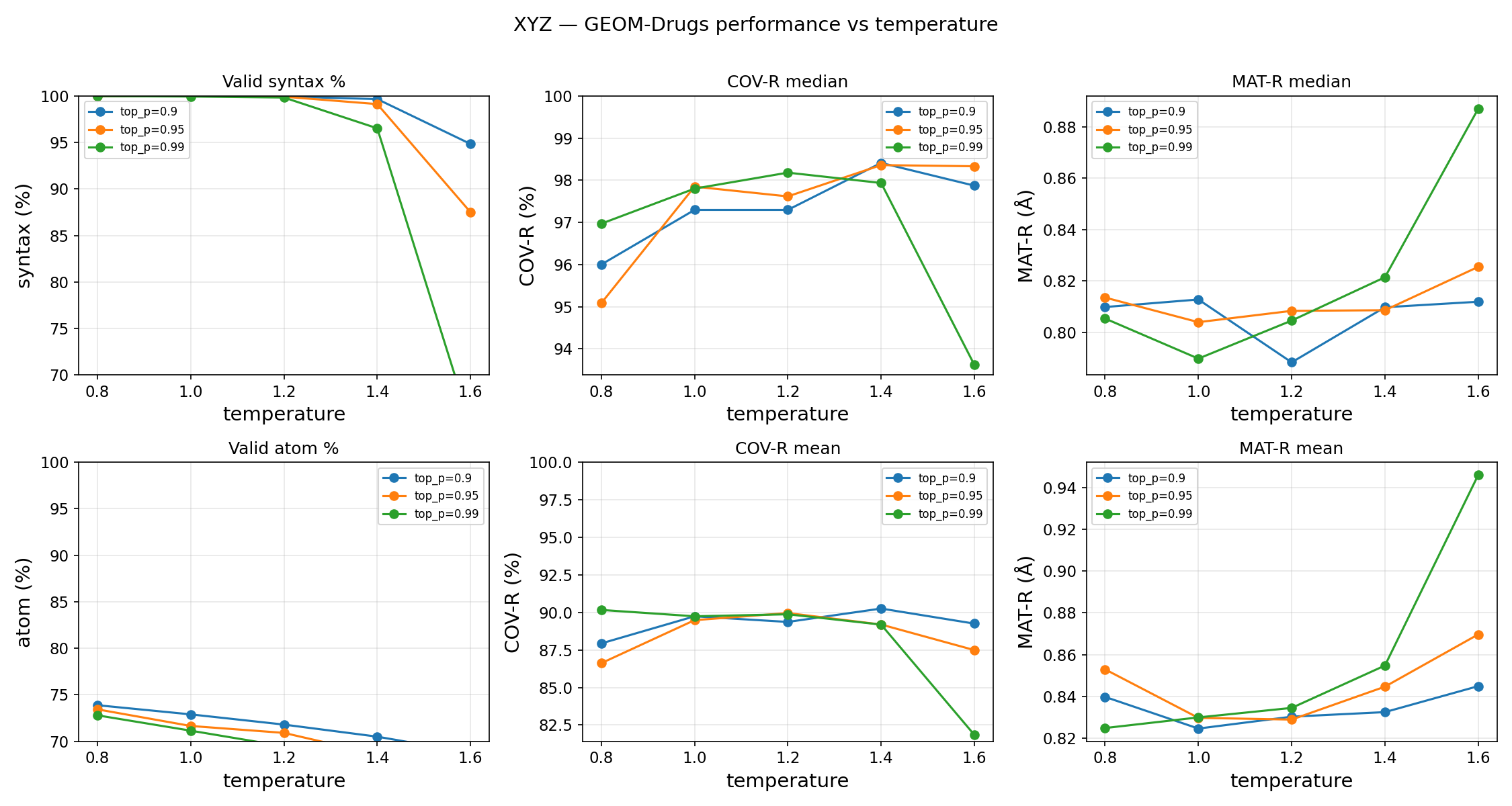}
\vspace{-2mm}
\caption{\textit{Temperature dependence of the GEOMLlama predictions on the GEOM-Drugs test data} for (top) Z-matrix and (bottom) XYZ formats. The optimal temperature for geometry prediction of GEOMLlama is higher than the training temperature of 1.0 and Top-P of 0.95. For Z-matrices the conformational sampling is found to be much better with negligible loss of validity, whereas validity plummets for the XYZ format due to poor atom counts.}
\label{fig:s3}
\end{figure}


\begin{figure}
\centering
\includegraphics[width=0.99\textwidth]{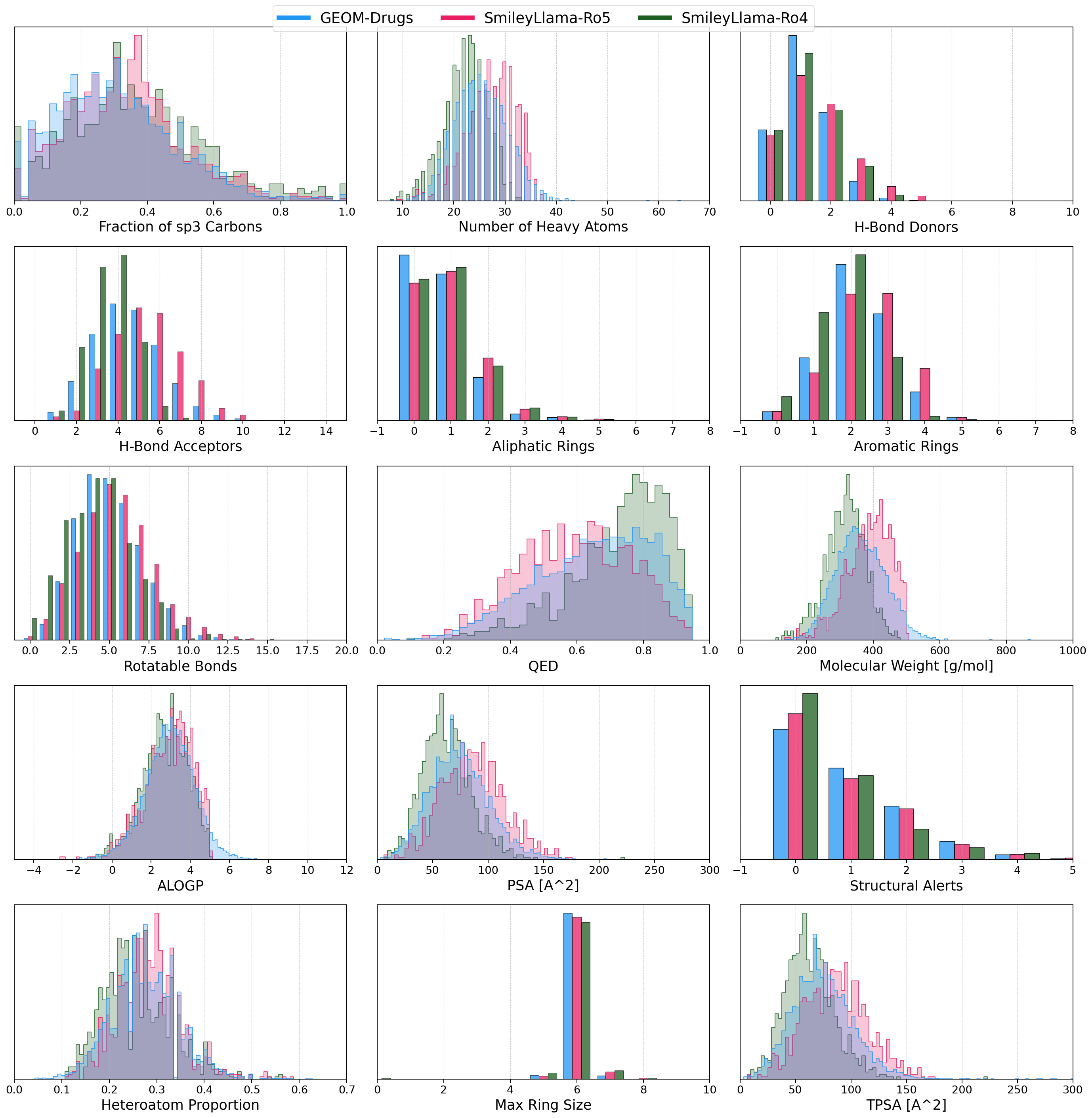}
\caption{\textit{Distribution comparisons for different properties of GEOM-Drugs training data (blue) with generated molecules from the test dataset from SmileyLlama Rule-of-Five (red) and SmileyLlama Rule-of-Four (green).} Overall the SmileyLlama Rule-of-Four is more similar to the Geom-Drugs training data.}
\label{fig:lstm_properties}
\end{figure}

\begin{figure}[H]
    \centering
        \includegraphics[width=0.95\linewidth]%
            {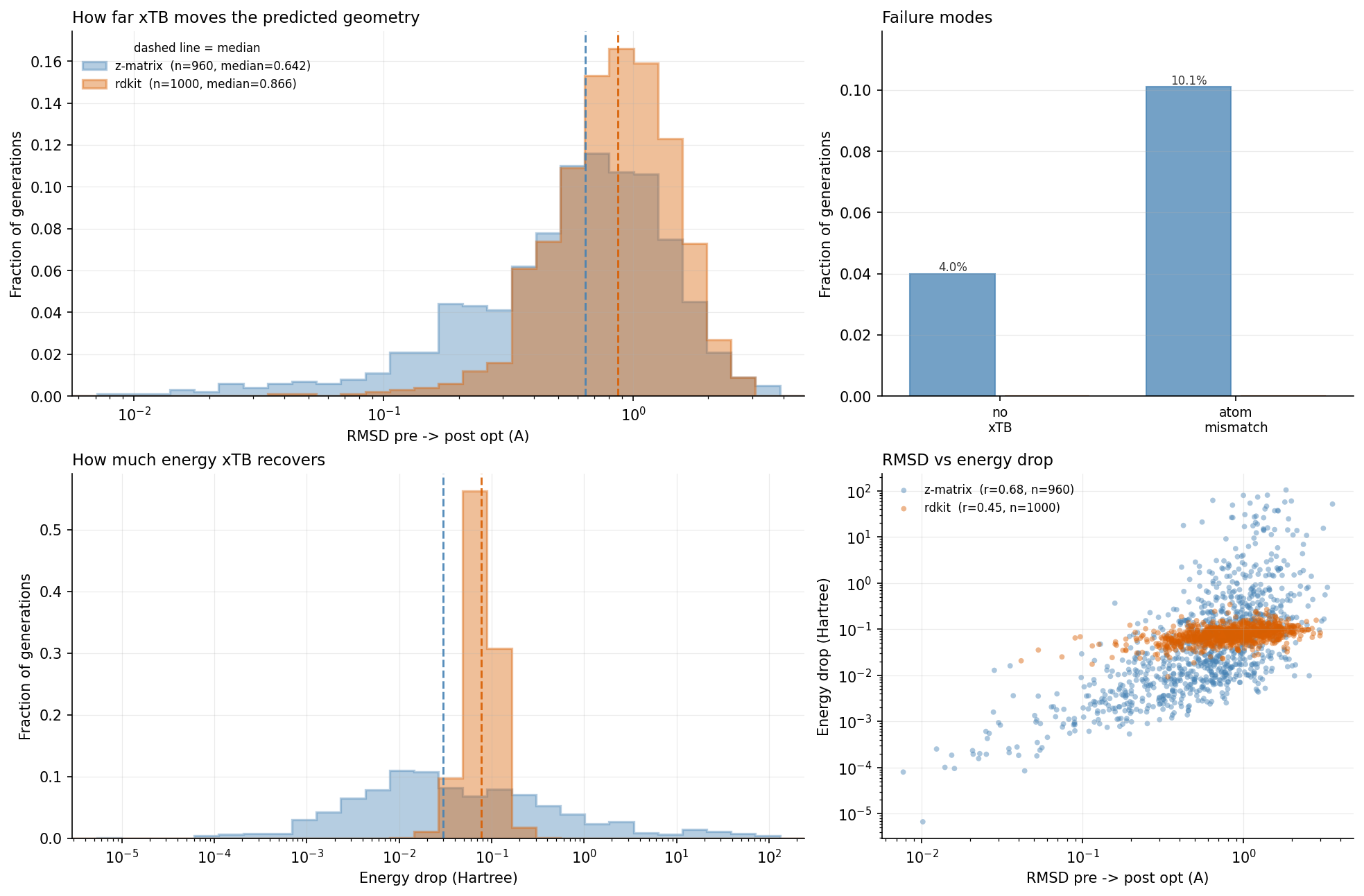}
    \caption{\textit{Performance of GeomLlama on SmileyLlama drug-like molecules prompted with Lipinski's Rule of Five.} (a) Representative geometries using GeomLlama trained with Z-matrices. (b) geometry differences between GeomLlama and RDKit's \texttt{EmbedMolecules} on SmileyLlama Rule-of-Five sequences compared against their xTB-optimized geometries. It is seen that GeomLlama has $\sim$8.4\% validity failures and a small set of optimization failures. (c) change in energy after optimizing GeomLlama's structures as well as RDKit's structures. (d) Change in geometry versus change in energy upon optimization with GFN2-xTB\cite{bannwarth_gfn2-xtbaccurate_2019}.}
    \label{fig:both_plots}
\end{figure}


\bibliography{library}